# Influence of the Hole Transporting Layer on the Thermal Stability of Inverted Organic Photovoltaics Using Accelerated Heat Lifetime Protocols

*Felix Hermerschmidt,[a] Achilleas Savva,[a] Efthymios Georgiou [a] , Sachetan M. Tuladhar,[b] James R. Durrant,[c] Iain McCulloch,[c] Donal D. C. Bradley,[d] Christoph J. Brabec,[e] Jenny Nelson,[b] Stelios A. Choulis[a,] \**

a Molecular Electronics and Photonics Research Unit, Department of Mechanical Engineering and Materials Science and Engineering, Cyprus University of Technology, 3041 Limassol, Cyprus

b Department of Physics, Imperial College London, London SW7 2AZ, UK

c Department of Chemistry, Imperial College London, London SW7 2AZ, UK

d Departments of Engineering Science and Physics, Division of Mathematical, Physical and Life Sciences, University of Oxford, Oxford OX1 3PD, UK

e Institute for Materials in Electronics and Energy Technology, Friedrich-Alexander University Erlangen-Nuremberg, 91054 Erlangen, Germany

ABSTRACT. High power conversion efficiency (PCE) inverted organic photovoltaics (OPVs) usually use thermally evaporated $MoO_3$ as a hole transporting layer (HTL). Despite the high PCE values reported, stability investigations are still limited and the exact degradation mechanisms of inverted OPVs using thermally evaporated $MoO_3$ HTL remain unclear under different environmental stress factors. In this study, we monitor the accelerated lifetime performance of non-encapsulated inverted OPVs using thiophene-based active layer materials and evaporated $MoO_3$ under the ISOS D-2 protocol (heat conditions 65 °C). The investigation of degradation mechanisms presented focus on optimized P3HT:PC[70]BM-based inverted OPVs. Specifically, we present a systematic study on the thermal stability of inverted P3HT:PC[70]BM OPVs using solution processed PEDOT:PSS and evaporated $MoO_3$ HTL. Using a series of measurements and reverse engineering methods we report that the P3HT:PC[70]BM/$MoO_3$ interface is the main origin of failure of P3HT:PC[70]BM-based inverted OPVs under intense heat conditions.

## 1. Introduction

Organic photovoltaics (OPVs) have attracted great scientific interest during the last decade due to their ease of manufacture with printable techniques and their potential to become flexible, lightweight and low cost energy sources.[1-2] High power conversion efficiencies and prolonged lifetimes are essential for OPV commercialization. Exciting power conversion efficiencies (PCEs) of 10% have been demonstrated,[3] but long stabilities of OPVs is the next barrier that needs to be overcome. Understanding the degradation mechanisms that influence the stability of different device configurations under various environmental stress factors is still a challenging task.

In the inverted structure, the use of a high work function metal such as silver results in enhanced lifetime compared to normal structured OPVs, which are limited in stability mainly due to oxidation of the low work function metals such as calcium or aluminum.[4] Poly(3,4-ethylenedioxythiophene) polystyrene sulfonate (PEDOT:PSS) is the most common material used as hole transporting layer (HTL) in both architectures. It is known that due to its hydroscopic nature, ingress of moisture and oxygen from the edges into the device can cause degradation.[5-6]

Furthermore, heat is one of the environmental factors found to significantly affect the long term stability of OPVs. Heat stability studies at operating temperature performed by Sachs-Quintana *et al.* on normal structure OPVs show that a thin polymer layer forms at the back (metal) contact, which creates an electron barrier between the active layer and the cathode.[7] This is related to the interface between the back contact and the bulk heterojunction and is shown to be the first step in thermal degradation in organic solar cells with normal architecture and especially while using low glass transition temperature conjugated polymers. However, in inverted OPVs this barrier formation is favorable as it helps hole selectivity. This feature has been claimed as an additional advantage of inverted OPVs.[7] Thus, the inverted structure is preferable concerning product development targets as it can allow more design flexibility and prolonged lifetime.

However, despite its prolonged lifetime, the top electrode (the anode) of inverted OPVs has been identified as one of the most vulnerable parts of the device under various environmental stress factors. Several studies were performed that reveal the crucial influence of the interface between the active layer and HTL in the lifetime performance of inverted OPVs.[8-11] Inverted OPVs with PEDOT:PSS treated with different wetting agents present different lifetime behavior in accelerated humidity conditions, thereby showing the crucial influence of processing and HTL/active layer interface formation.[12] Furthermore, Norrman *et al.* have suggested that PEDOT:PSS is the main factor of degradation of inverted OPVs in ambient conditions. They have demonstrated that the phase separation of PEDOT:PSS into PEDOT rich and PSS rich regions and the active layer/PEDOT:PSS interface are the main sources of failure of the long-term lifetime of inverted OPVs.[13]

Therefore, the substitution of the moisture sensitive PEDOT:PSS in inverted OPVs with metal oxide HTLs such as $MoO_3$, $V_2O_5$, and $WO_3$ could be beneficial in some cases concerning stability of inverted OPVs. However, the use of solution processed metal oxides as HTL in inverted OPVs is still challenging, since major processing issues have to be addressed. Recent work has shown that solution-processed $MoO_3$ can be used to produce efficient and stable normal structure and inverted OPVs when used as a sole HTL, [14-15] as well as when inserted in addition to PEDOT:PSS.[16] Mixing of solution-based $MoO_3$ with PEDOT:PSS has also yielded efficient inverted solar cells.[17] However, evaporated metal oxides are still more widely used as HTLs in inverted OPVs, leading to high PCEs with optimum hole selectivity.

Evaporated $MoO_3$ is a promising material as an HTL in inverted OPVs in terms of efficiency due to the favorable energy level alignment between its work function measured at $6.7eV$,[18] and the HOMO of typical active layers in the range of -5.0 eV for polythiophene-based materials.[19] There are promising developments regarding the stability of inverted OPVs comparing $MoO_3$ and PEDOT:PSS HTL in air,[15] however lifetime studies under humidity comparing $MoO_3$ and PEDOT:PSS HTL show that evaporated $MoO_3$ is also very sensitive to moisture and leads to inverted OPVs with limited lifetime.[20] The reason is that oxygen and humidity ingress can alter the conductivity and work function of $MoO_3$.[21-22] On the other hand, the hygroscopic PEDOT:PSS in this architecture can act as a getter and protect the active layer from water interactions.[23] Furthermore, $MoO_3$ may undergo a change in the oxidation state of the Mo (VI) metal centre under exposure to heat or UV light, leading to changes in its work function and its optical response. Under heat conditions, $MoO_3$ releases oxygen leading to lower Mo oxidation states and a shift in the work function of the oxide.[22, 24]

In addition, Voroshazi et al. have shown that the reduction of $MoO_3$ can also occur under heat in dark conditions in inverted OPVs with $MoO_3$/Ag/Al top contact due to chemical interactions between $MoO_3$ and Al atoms.[25] Another aspect of the degradation mechanism of inverted OPVs with the aforementioned top electrode was pointed out by Rösch et al. through different imaging techniques, whereby they attributed the failure to the migration of silver and diffusion into the $MoO_3$ layer, changing its work function.[26] However, a full understanding of the stability of $MoO_3$ HTL in inverted OPVs is still lacking due to the complexity of this material and the varying environmental stress factors that can influence the degradation cause of inverted OPVs with $MoO_3$ HTL.

In this study, therefore, we aim to analyze the degradation mechanisms of non-encapsulated inverted OPVs using $MoO_3$ as HTL, focusing on accelerated heat lifetime conditions using the ISOS-D-2 protocol (dark – low RH – 65 °C). As a first step, inverted OPVs using various polymer:fullerene active layers are investigated. Photocurrent mapping measurements provide degradation images at 65 °C at various points in time. Secondly, different electrode configurations are investigated for our reference P3HT:PC[70]BM polymer-fullerene active layer. Using combinations of two interlayers – $MoO_3$ and PEDOT:PSS – at the top electrode of P3HT:PC[70]BM based devices, the device degradation is attributed primarily to the interface between the active layer and $MoO_3$ and secondly to that between $MoO_3$ and Ag. Thirdly, alternative reverse engineering methods are used to enhance our assumption that the $MoO_3$/Ag interface significantly contributes to the degradation of devices under heat.

## 2. Experimental

Pre-patterned glass-ITO substrates (sheet resistance 4 Ω/sq) were purchased from Psiotec Ltd. Zinc acetate dehydrate, 2-methoxyethanol and ethanolamine were purchased from Sigma Aldrich, P3HT from Rieke Metals, PTB7 from 1-Material, PC[70]BM from Solenne BV, PEDOT:PSS PH from H.C. Stark, $MoO_3$ powder from Sigma Aldrich and silver pellets from Kurt J. Lesker.

For the fabrication of inverted solar cells, ITO substrates were sonicated in acetone and subsequently in isopropanol for 10 minutes. ZnO electron transporting layer was prepared using a sol-gel process as described in detail elsewhere.[27] The ZnO precursor was doctor bladed on top of ITO substrates and annealed for 20 min at 140 °C in ambient conditions. After the annealing step a 40 nm ZnO layer is formed as measured with a Veeco Dektak 150 profilometer. The photo-active layers, which are blends of 36 mg/ml in chlorobenzene P3HT:PC[70]BM (1:0.8 by weight), 15 mg/ml in o-DCB DPPTTT:PC[70]BM (1:2 by weight), 25 mg/ml PTB7:PC[70]BM (1:1.5 by weight) were doctor bladed on top of ZnO resulting in a layer thickness of ~180 nm, 100 nm and 90 nm, respectively, as measured with a profilometer.

For all PEDOT:PSS-based devices a treatment with two wetting agents (Zonyl and Dynol) has been applied as described in detail previously,[28] and will be named as PEDOT:PSS:ZD within the text and figures. The P3HT:PC[70]BM-based inverted OPVs were annealed inside a glovebox at 140°C for 20 minutes prior to thermal evaporation of a silver layer with a thickness of 100 nm, resulting in four solar cells, each with an active area of 9 mm$^2$. For the devices with MoO$_3$ HTL, 10 nm of MoO$_3$ was evaporated prior to silver evaporation. For devices using MoO$_3$ and PEDOT:PSS:ZD as HTL, MoO$_3$ was evaporated onto the active layer, PEDOT:PSS:ZD doctor bladed onto the MoO$_3$ layer, followed by silver evaporation. These devices were annealed for another 2 min at 140 °C after PEDOT:PSS:ZD deposition. For devices with PEDOT:PSS:ZD/MoO$_3$/Ag top electrode the procedure was the same as for PEDOT:PSS:ZD-based devices, except for an extra 10nm MoO$_3$ layer evaporated on top of the PEDOT:PSS:ZD layer.

The current density-voltage (J/V) characteristics were measured with a Keithley source measurement unit (SMU 2420). For illumination, a calibrated Newport Solar simulator equipped with a Xe lamp was used, providing an AM1.5G spectrum at 100mW/cm$^2$ as measured by an Oriel 91150V calibration cell equipped with a KG5 filter. Photocurrent measurements were performed under 405 nm laser excitation using a Botest PCT photocurrent system. Thermal aging of the devices has been performed according to ISOS-D-2 protocol,[29] namely, inverted OPVs under test were placed in a dark chamber (Binder) at 65 °C with controlled ambient humidity ~45% and aged for several hours. Their J/V characteristics were measured at different stages of degradation. The devices were un-encapsulated and the lifetime study was started 1 day after device fabrication.

## 3. Results and Discussion

From our initial heat lifetime experiments comparing inverted OPVs comprised of ITO/ZnO/P3HT:PC[70]BM with a $MoO_3$/Ag vs PEDOT:PSS:ZD/Ag top electrode, we observed a dramatic drop in all photovoltaic parameters after just a few hours of aging of the inverted OPVs using $MoO_3$ as HTL (data not shown).

### 3.1. Investigating the degradation mechanisms of inverted OPVs under the ISOS-D-2 protocol using different polymer:fullerene blends and top electrode configurations

In order to obtain a general picture about the heat degradation origins of inverted OPVs, three different polymer:fullerene blends were tested. The inverted OPVs with the following general configuration: ITO/ZnO/polymer:PC[70]BM/$MoO_3$/Ag or PEDOT:PSS:ZD/Ag were exposed to heat conditions based on the ISOS-D-2 protocol (T = 65 °C, dark, RH = constant ambient ~40%). P3HT:PC[70]BM (1:0.8),[27] PTB7:PC[70]BM (1:1.5),[30] and DPPTTT:PC[70]BM (1:4)[33] were used as the active layers. The polymer:fullerene ratios used were chosen from the best performing devices reported in literature. The average absolute values of the photovoltaic parameters for the inverted OPVs under study were obtained just before starting the heat aging and are summarized in Table 1.

**Table 1.** Average absolute photovoltaic parameter values out of 8 inverted OPVs in each case, obtained before initiating the heat aging.

| Inverted OPVs | Voc (V) | Jsc (mA/cm²) | FF (%) | PCE (%) |
|---|---|---|---|---|
| ITO/ZnO/P3HT:PC[70]BM/$MoO_3$/Ag | 0.56 | 10.42 | 58.9 | 3.54 |
| ITO/ZnO/PTB7:PC[70]BM/$MoO_3$/Ag | 0.68 | 12.49 | 46.8 | 3.97 |
| ITO/ZnO/DPPTTT:PC[70]BM/$MoO_3$/Ag | 0.55 | 11.22 | 52.9 | 3.32 |

| | | | |
|---|---|---|---|
| ITO/ZnO/P3HT:PC[70]BM/PEDOT:PSS:ZD/Ag | 0.58 | 9.26 | 58.3 | 3.16 |
| ITO/ZnO/PTB7:PC[70]BM/PEDOT:PSS:ZD/Ag | 0.69 | 8.26 | 50.3 | 2.86 |
| ITO/ZnO/DPPTTT:PC[70]BM/PEDOT:PSS:ZD/Ag | 0.54 | 8.07 | 54.3 | 2.53 |

It should be noted that while some of the device performance values presented in Table 1 may be below some literature values for the device structure, the purpose of this paper is not to produce best-performing "hero" devices. Thus the relative performance values can still be used to obtain a general picture for the degradation behavior. Figure 1 shows the normalized photovoltaic parameters over time of exposure under heat conditions of all the inverted OPVs under study.

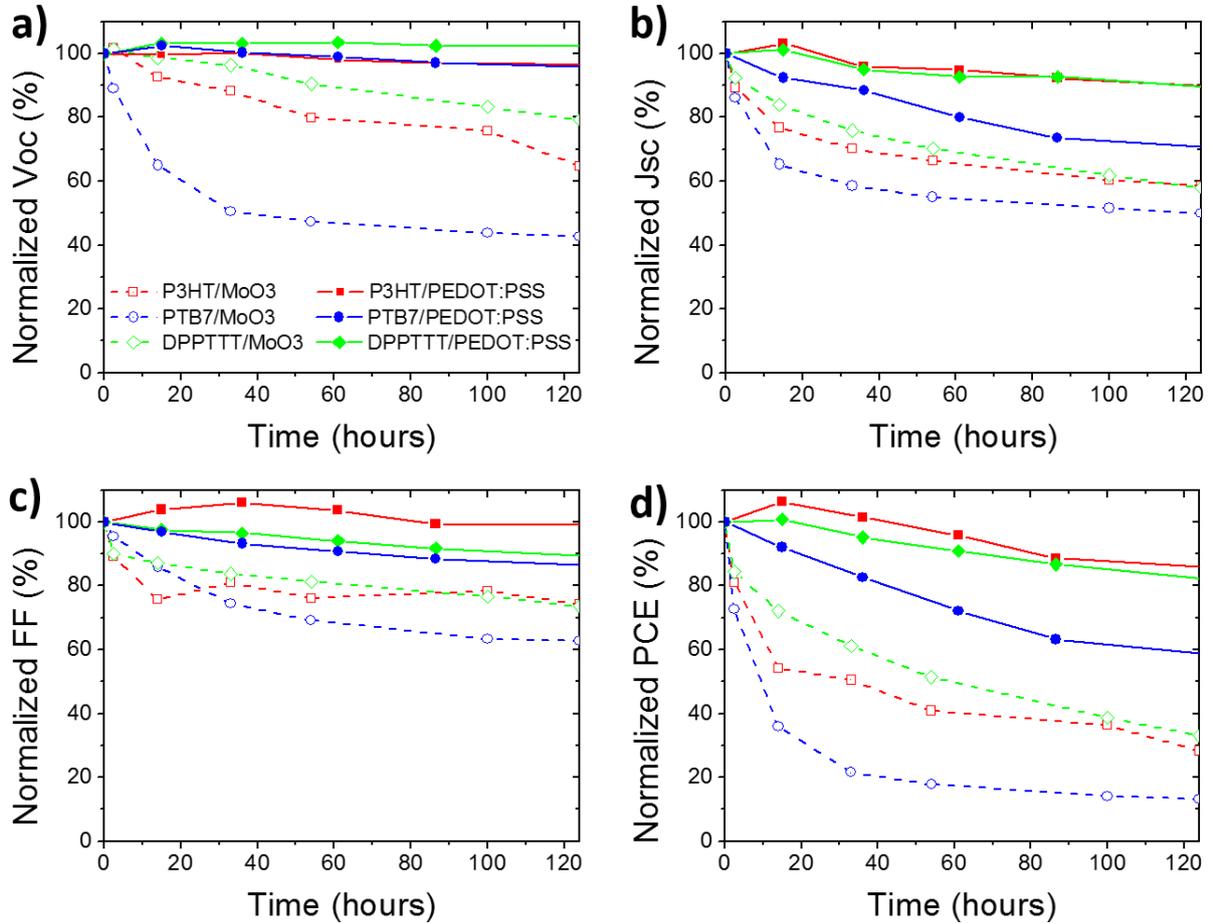

**Figure 1.** Degradation trends of the OPV parameters at 65 °C over time for non-encapsulated inverted OPVs with different active layers (red P3HT:PC[70]BM, green DPPTTT:PC[70]BM, blue PTB7:PC[70]BM) using either MoO3 (dashed lines) or PEDOT:PSS (solid lines) as hole transporting

layers, plotted as a function of (a) normalized Voc, (b) normalized Jsc, (c) normalized FF and (d) normalized PCE.

As shown in Figure 1 all the inverted OPVs with different polymer:fullerene blends using MoO$_3$/Ag top electrodes drop dramatically within the first few hours of exposure under accelerated heat conditions. In contrast, the lifetime performance of the corresponding inverted OPVs using PEDOT:PSS HTL is significantly better to that observed with MoO$_3$ for all the different polymer:fullerene blend combinations studied.

It is worth noting that all the polymers used in this study are thiophene-based (P3HT, PTB7 and DPPTTT). For PTB7:PC[70]BM-based inverted OPVs, a fast degradation pattern under 85 °C dark heat has previously been observed and been primarily attributed to morphological changes within the active layer.[31] Furthermore, PTB7 has been reported as sensitive to several environmental factors,[32-33] while DPPTTT has shown greater stability under light and oxygen,[34] but not under thermal conditions.[35] On the other hand, P3HT:PC[70]BM-based inverted OPVs are known to be resistive to several environmental stress factors and impressive lifetime performances in accelerated and outdoor conditions are demonstrated elsewhere.[36-37]

All the above observations indicate that the incorporation of MoO$_3$ as HTL at the top electrode of inverted OPVs has a dominant influence on the lifetime performance when subjected to the ISOS D-2 protocol despite the complicated degradation patterns that might also arise due to the variety of active layers. Therefore, our further analyses in the remainder of this study focus on the understanding of the degradation mechanisms of inverted OPVs using our reference and well optimized P3HT:PC[70]BM as the active layer.

## 3.2. Investigating the degradation mechanisms of P3HT:PC[70]BM/MoO₃-based inverted OPVs under the ISOS-D-2 protocol using buffer layer engineering.

In order to visualize this fast degradation of P3HT:PC[70]BM-based solar cells, photocurrent images that map the device active area and its degradation are shown in **Error! Reference source not found.**. Inverted OPVs using P3HT:PC[70]BM as active layer and MoO₃ as HTL were aged and photocurrent  mapping images were extracted at different stages of degradation. The degradation of inverted OPVs with MoO₃ as HTL happens very fast, even after 5 hours degradation under heat, and the photocurrent drops dramatically (see Figure 1 and **Error! Reference source not found.**).

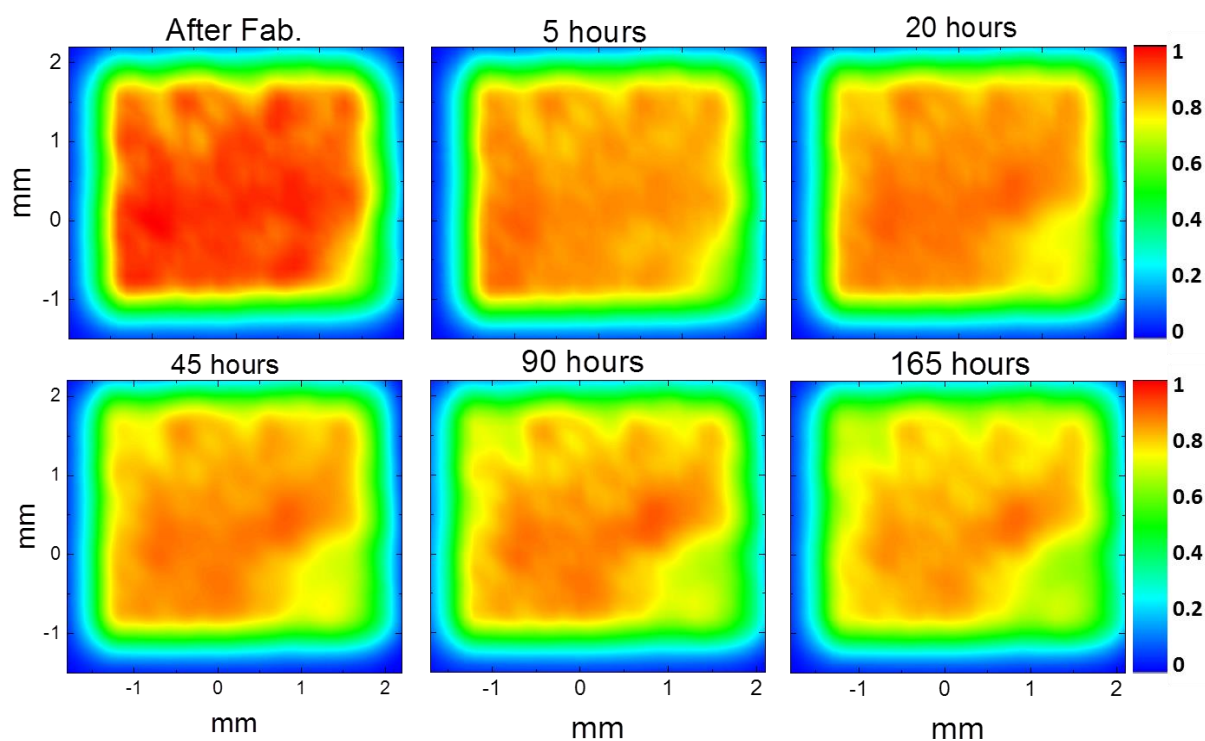

**Figure 2.** Normalized photocurrent mapping images of non-encapsulated inverted OPVs using MoO₃ as HTL, showing degradation at 65 °C over time of exposure.

The photocurrent mapping measurements indicate that for non-encapsulated inverted P3HT:PC[70]BM solar cells incorporating evaporated MoO₃ HTL, the photocurrent generation drops dramatically in just 5 hours of exposure. The photocurrent intensity observed is significantly less within the whole device area. Despite the obvious drop near the edges of the devices which could be attributed to ingress of air and moisture, the center of the device also shows significantly less photocurrent generation compared to the fresh devices. This is in accordance to Figure 1b which shows a fast Jsc degradation of P3HT:PC[70]BM/MoO$_3$ within the first few hours of exposure.

To investigate the interfacial interaction of MoO$_3$ with the P3HT:PC[70]BM active layer, we apply buffer layer engineering to isolate the interfaces under study. Inverted OPVs based on P3HT:PC[70]BM with four different configurations of top electrode – MoO$_3$/Ag, PEDOT:PSS:ZD/Ag, PEDOT:PSS:ZD/MoO$_3$/Ag and MoO$_3$/PEDOT:PSS:ZD/Ag – were tested under 65 °C. The photovoltaic parameters of these devices are shown in Table 2.

**Table 2.** Initial photovoltaic parameters of inverted OPVs with different hole transporting layers.

| Inverted P3HT:PC[70]BM OPVs using different top electrode configurations | Voc (V) | Jsc (mA/cm$^2$) | FF (%) | PCE (%) |
|---|---|---|---|---|
| PEDOT:PSS:ZD/Ag | 0.58 | 9.85 | 54.46 | 3.13 |
| PEDOT:PSS:ZD/MoO$_3$/Ag | 0.57 | 10.03 | 53.65 | 3.08 |
| MoO$_3$/Ag | 0.56 | 11.17 | 56.41 | 3.54 |
| MoO$_3$/PEDOT:PSS:ZD/Ag | 0.55 | 8.52 | 57.20 | 2.72 |

We observe that the solar cells using MoO$_3$/Ag top electrode show the highest initial efficiency of 3.54%. This is followed by very similar device efficiencies obtained using PEDOT:PSS:ZD/Ag and PEDOT:PSS:ZD/MoO$_3$/Ag of 3.13% and 3.08%, respectively. Finally, 2.72% PCE is obtained for inverted OPVs with an MoO$_3$/PEDOT:PSS:ZD/Ag top electrode. Interestingly, these values are comparable to those reported by Wang *et al.* using a solution-based hybrid electrode consisting of MoO$_3$ and PEDOT:PSS,[17] indicating an efficient bilayer in our system.

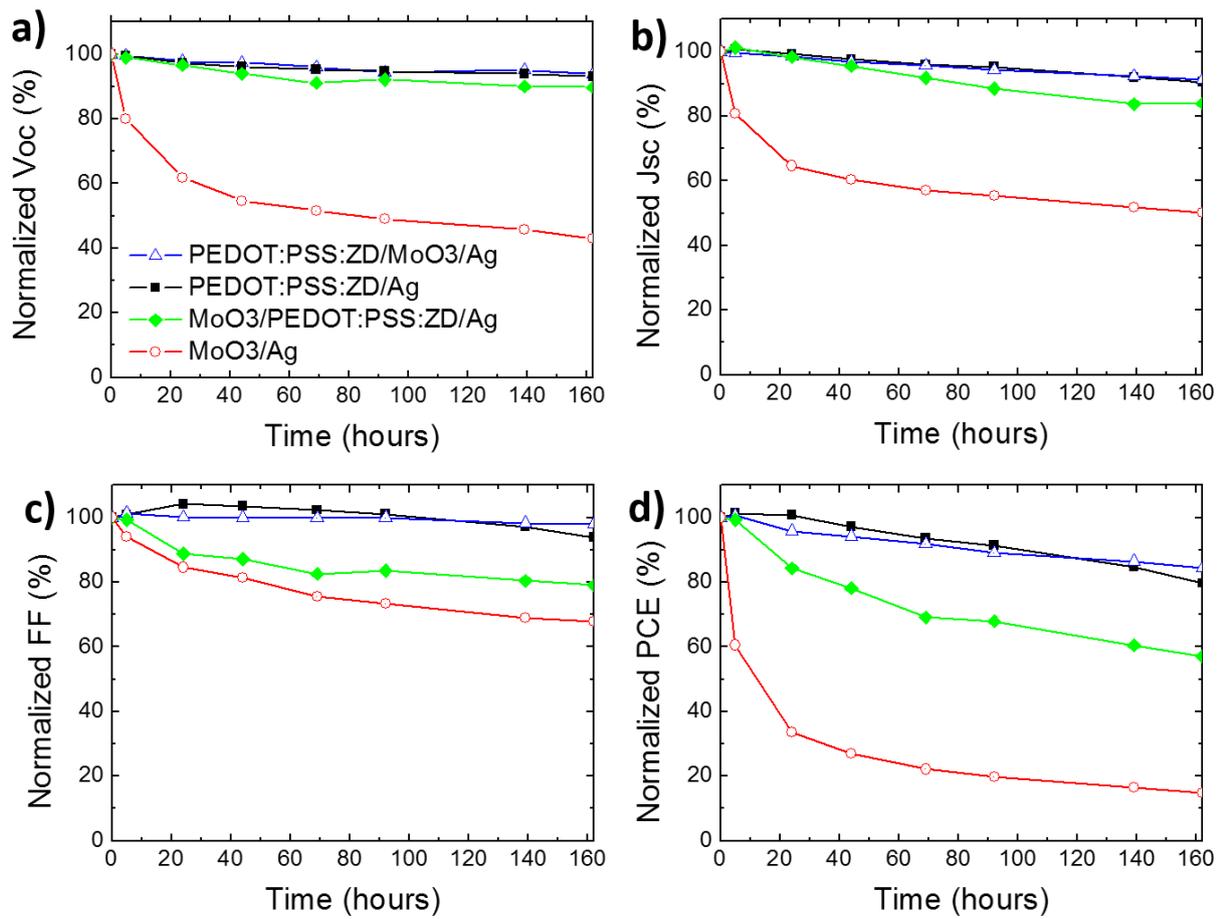

**Figure 3.** Degradation trends of inverted OPVs parameters at 65 °C over time for ITO/ZnO/P3HT:PC[70]BM with different top electrode configurations as a function of (a) normalized Voc, (b) normalized Jsc, (c) normalized FF and (d) normalized PCE.

As shown in

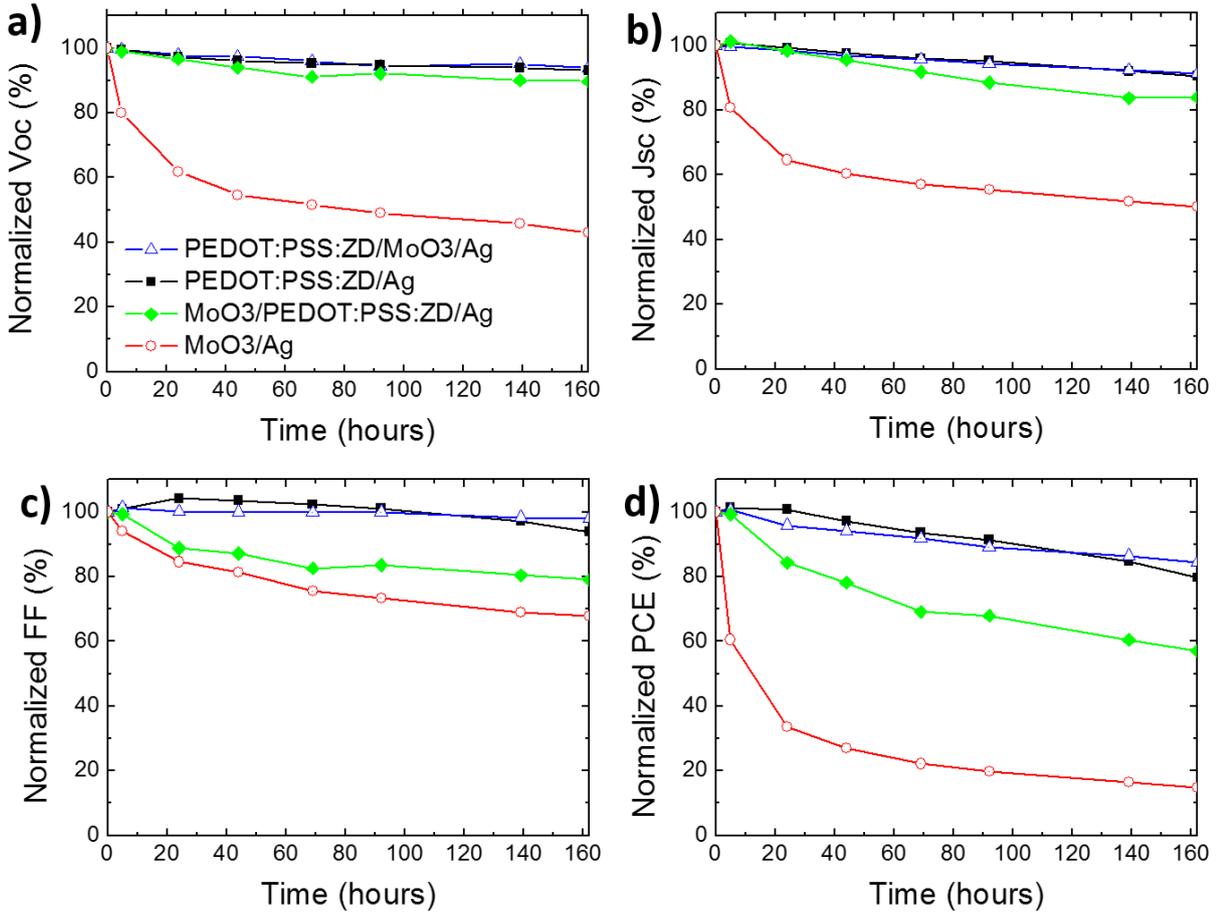

Figure **3** and in line with the previous observations from Figure 1, inverted OPVs using MoO₃/Ag top electrode degrade very fast under heat conditions from the initial efficiency of 3.54% compared to inverted OPVs with PEDOT:PSS:ZD as HTL. As also reported elsewhere,[38] all the J/V parameters exponentially drop without allowing precise identification of the failure mechanism. However, by incorporating a PEDOT:PSS:ZD layer between the active layer and MoO₃ as well as between MoO₃ and Ag, the degradation behavior can be influenced.

The inverted OPVs with MoO₃/PEDOT:PSS:ZD/Ag as a top contact present an intermediate PCE decay between devices with MoO₃ and PEDOT:PSS:ZD as HTL. The majority of the PCE drop on those devices is attributed to FF (

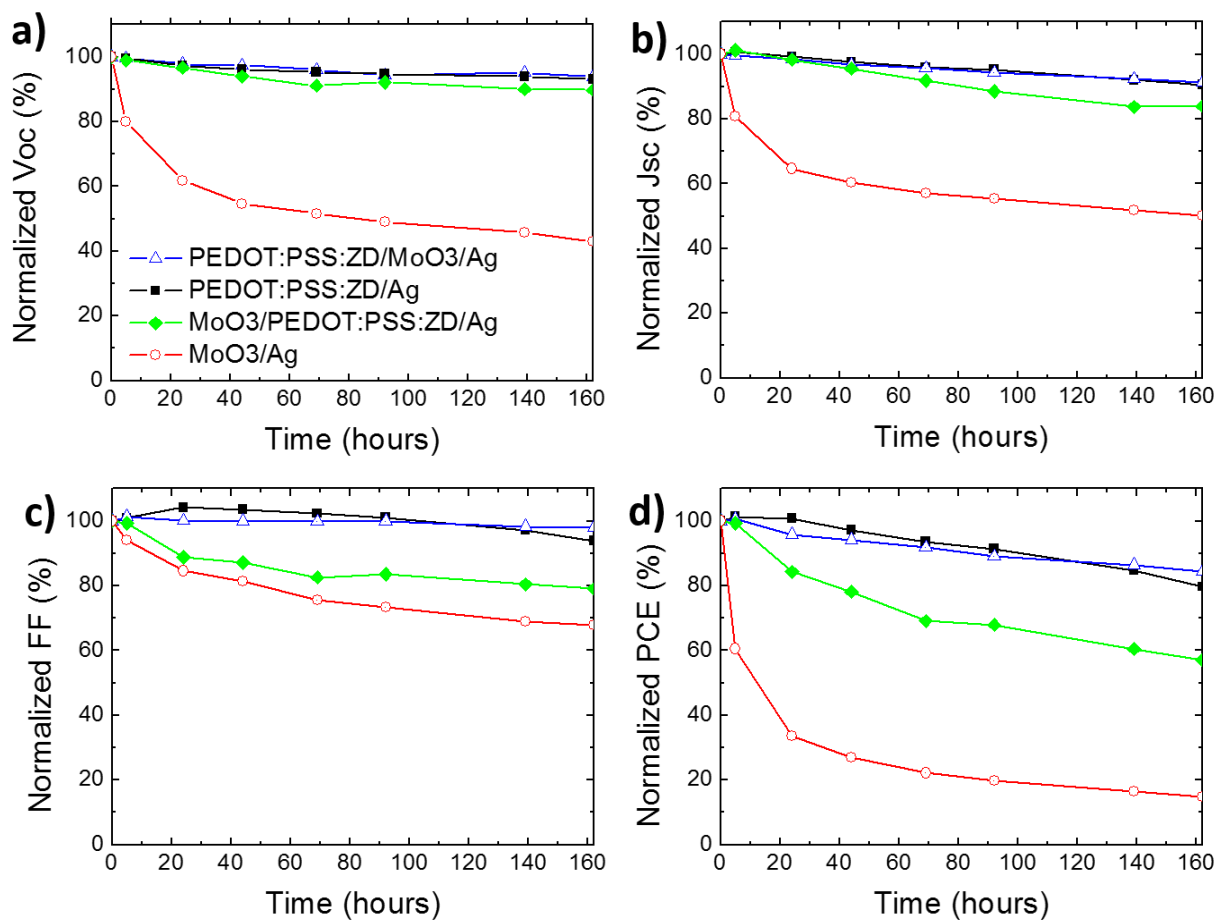

Figure **3**). Thus, we assume that the influence on FF in devices with MoO₃/Ag top contact is attributed to the interface of MoO₃ and active layer. By inserting the layer of PEDOT:PSS:ZD between MoO₃ and Ag, the suppressed decay of those devices indicates that direct contact between the MoO₃ and Ag influences device degradation. Furthermore, the Voc decline of devices with a direct interface between P3HT:PC[70]BM and MoO₃ (organic/metal oxide interface) is more obvious than the devices with P3HT:PC[70]BM and PEDOT:PSS:ZD (organic/organic interface). The latter can again link the failure of the device to the interface of the P3HT:PC[70]BM active layer with MoO₃, as has been reported previously.[38]

On the other hand inverted OPVs with PEDOT:PSS:ZD/MoO$_3$ HTL present similar lifetime behavior to inverted OPVs with only PEDOT:PSS:ZD as HTL. In that case no obvious degradation is observed when MoO$_3$ and Ag are in contact. Two possible phenomena could explain this.

The first is that carrier selectivity occurs through PEDOT:PSS:ZD and carrier (i.e. hole) collection through MoO$_3$/Ag , suggesting that electrons are responsible for the degradation at the MoO$_3$ / Ag interface and that degradation is prevented when the MoO$_3$ / Ag is protected from electrons. This effect has been seen in work by Zhu *et al.* on inverted hybrid quantum dot/polymer solar cells,[39] in which use of a dual HTL (PEDOT:PSS/MoO$_3$) led to enhanced electron blocking. This could explain the reduced degradation characteristics observed in our OPVs in which PEDOT:PSS interfaces with the active layer.

The second might be that silver migration and diffusion in the active layer through defects of the MoO$_3$ layer is prevented by inserting the PEDOT:PSS:ZD layer, as assumed elsewhere.[26, 38] This effect of atoms diffusing into the active layer has been investigated in several works: The Wantz group has observed diffusion of silver atoms in MoO$_3$ and organic layers, but, interestingly, they did not observe oxygen migration.[40] Additionally, the Österbacka group saw that diffusion of molecules into the active layer material causes doping effects that are detrimental to device performance.[41] Our findings for the MoO$_3$/Ag top contact seem to reflect these behaviors. However, since upon insertion of the PEDOT:PSS:ZD layer in our inverted OPVs the detrimental effects are slowed down drastically, we believe this is a strong indication that indeed the silver top contact also influences these degradation effects.

Therefore, in our experimental approach, our observations reveal that the interfaces between the active layer and MoO$_3$ as well as between MoO$_3$ and Ag majorly contribute to the degradation of the P3HT:PC[70]BM inverted OPVs under heat aging conditions, and that both electron blocking as well as diffusion of both MoO$_3$ and Ag species may play a role in this.

To better understand the above observations on the degradation mechanism, the dark J/V characteristics of the different types of OPVs with different buffer layers at the initial and at different stages of degradation were obtained and are shown in

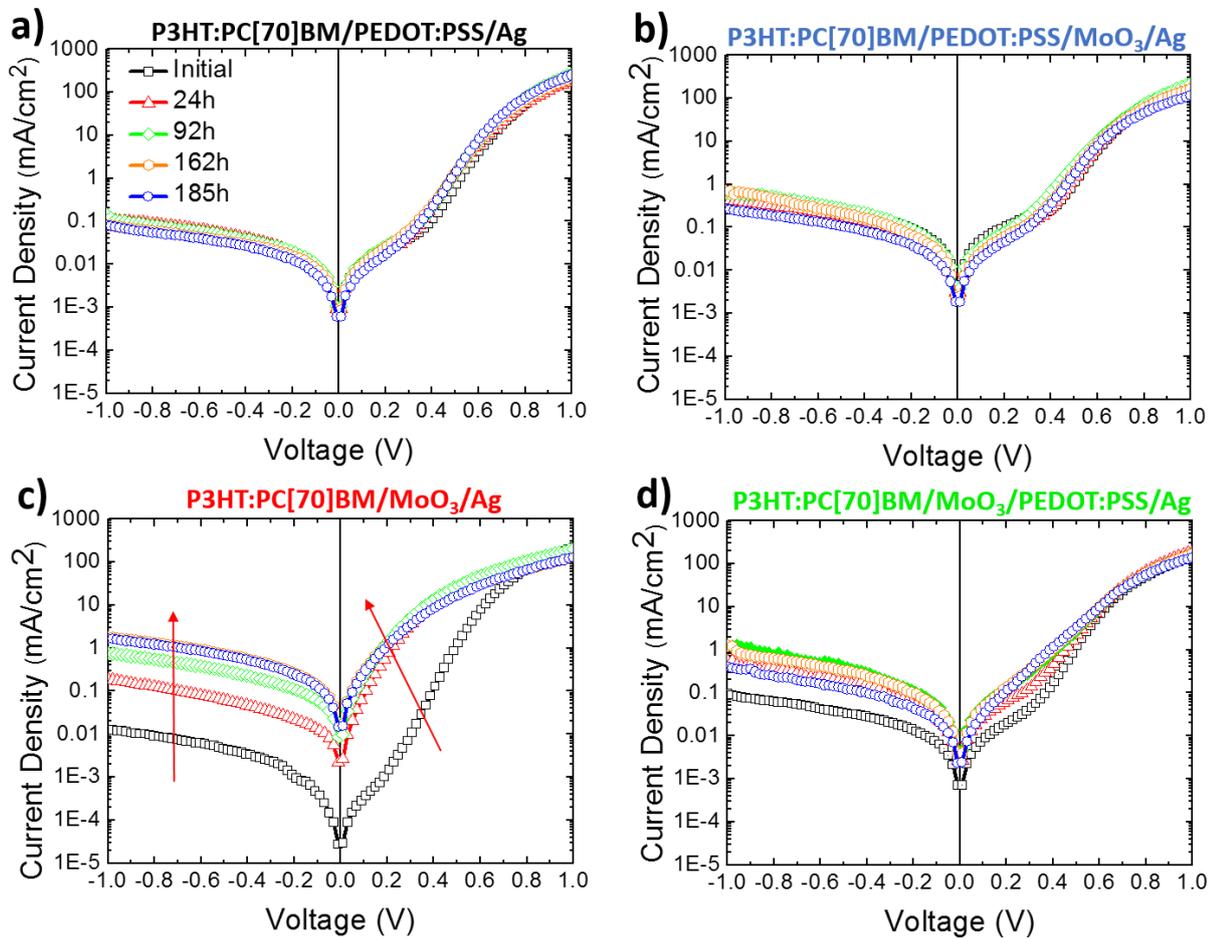

Figure **4**.

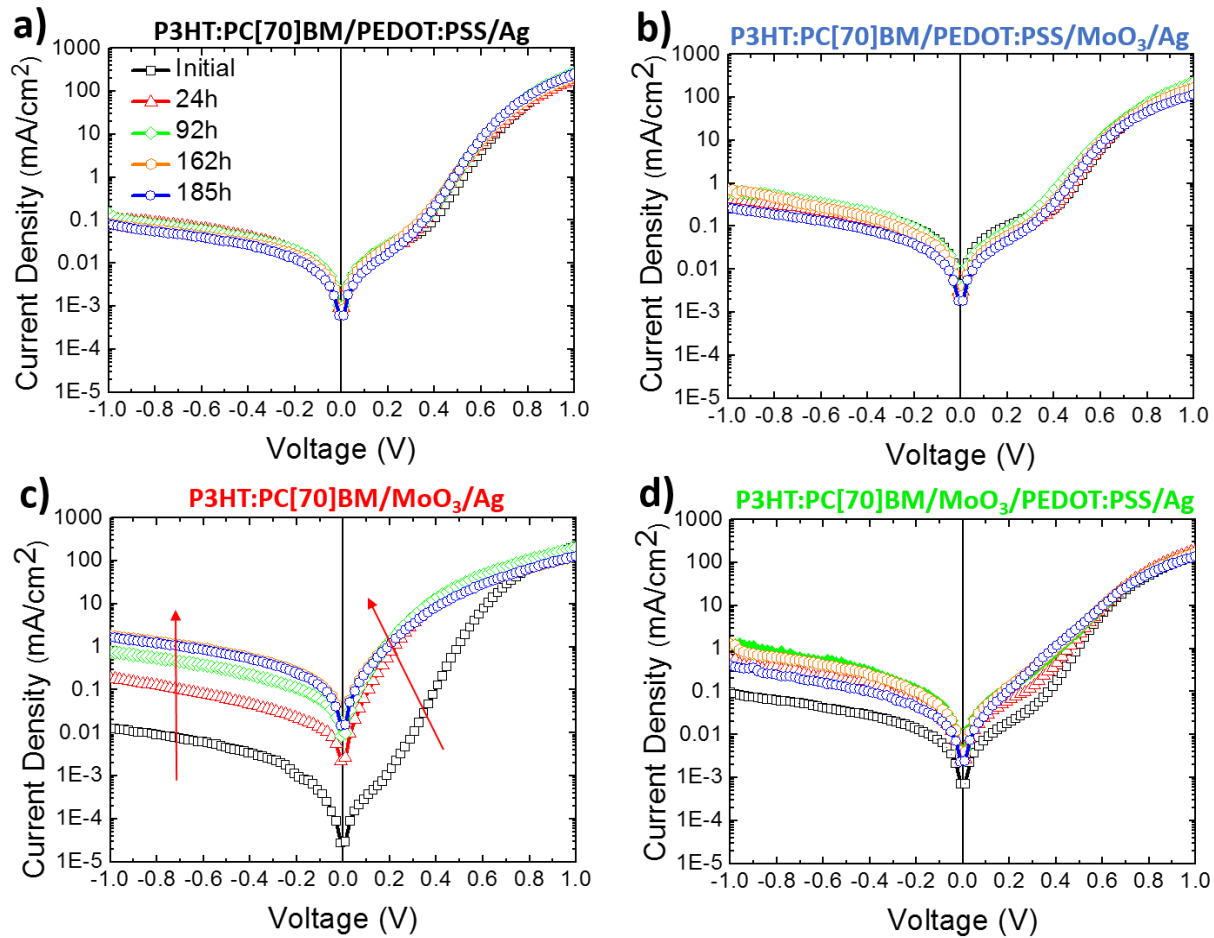

**Figure 4.** Current density versus voltage characteristics (J/V) in dark over time of exposure under heat conditions for inverted OPVs using different hole transporting layers – (a) PEDOT:PSS:ZD, (b) PEDOT:PSS:ZD/MoO₃, (c) MoO₃ and (d) MoO₃/ PEDOT:PSS:ZD.

The inverted OPVs containing MoO₃ HTL as produced and measured in the dark (see Figure 4) have low leakage current at reverse bias implying high shunt resistance (Rp). In addition at high forward bias they have high current density and thus low series resistance (Rs) value and therefore their diode-like behavior is better than PEDOT:PSS:ZD-containing devices and accordingly they present higher PCE values, as shown in Table 1.

The dark J/V curves of devices with PEDOT:PSS:ZD as HTL show more stable Rs and Rp values, which are slightly improved over time (

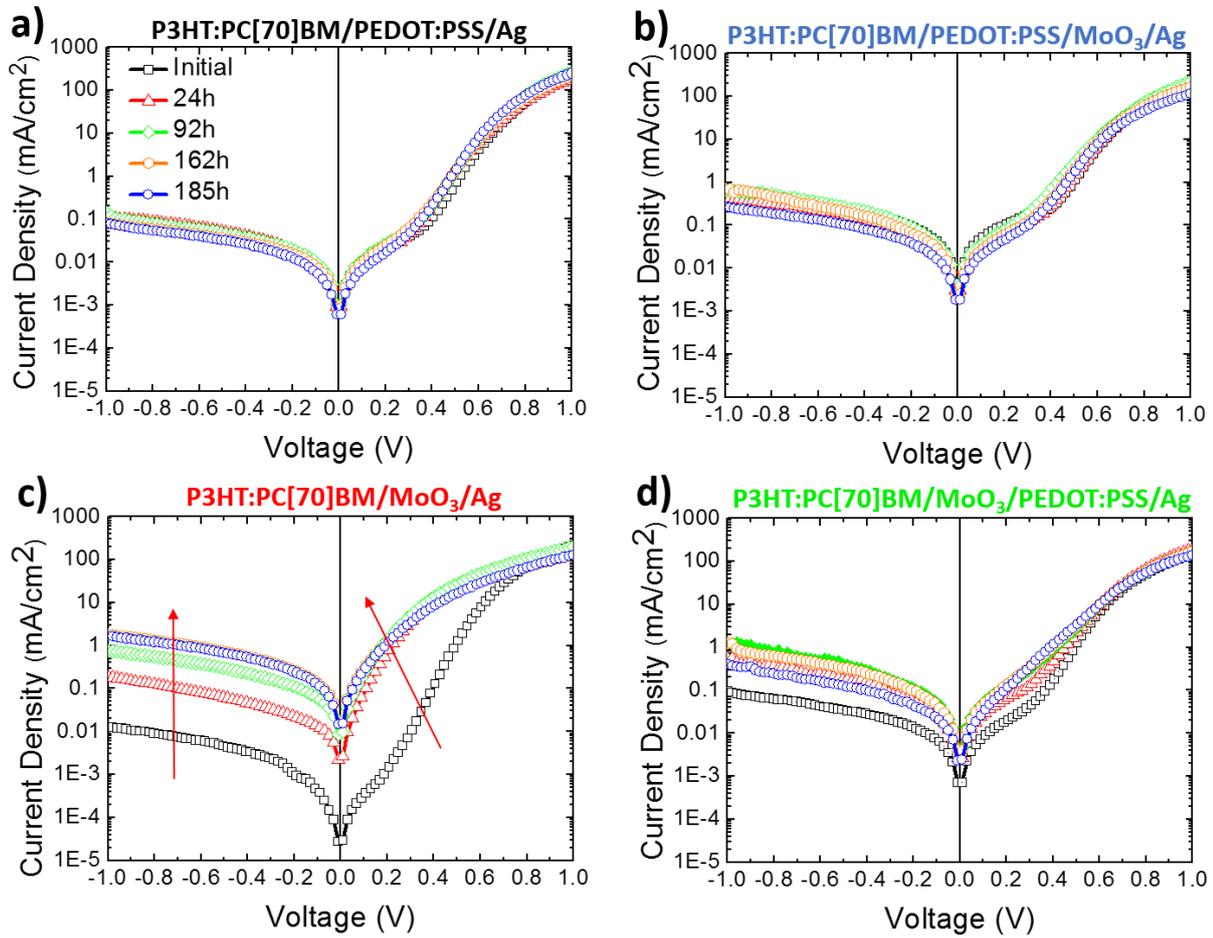

Figure **4**a). This shows that the organic-organic interface improves over time under heat conditions and that hole transportation is favored at the first stages of degradation. In addition, the performance may be assisted by formation of a hole selective thin P3HT layer on top of the active layer under heating [7].

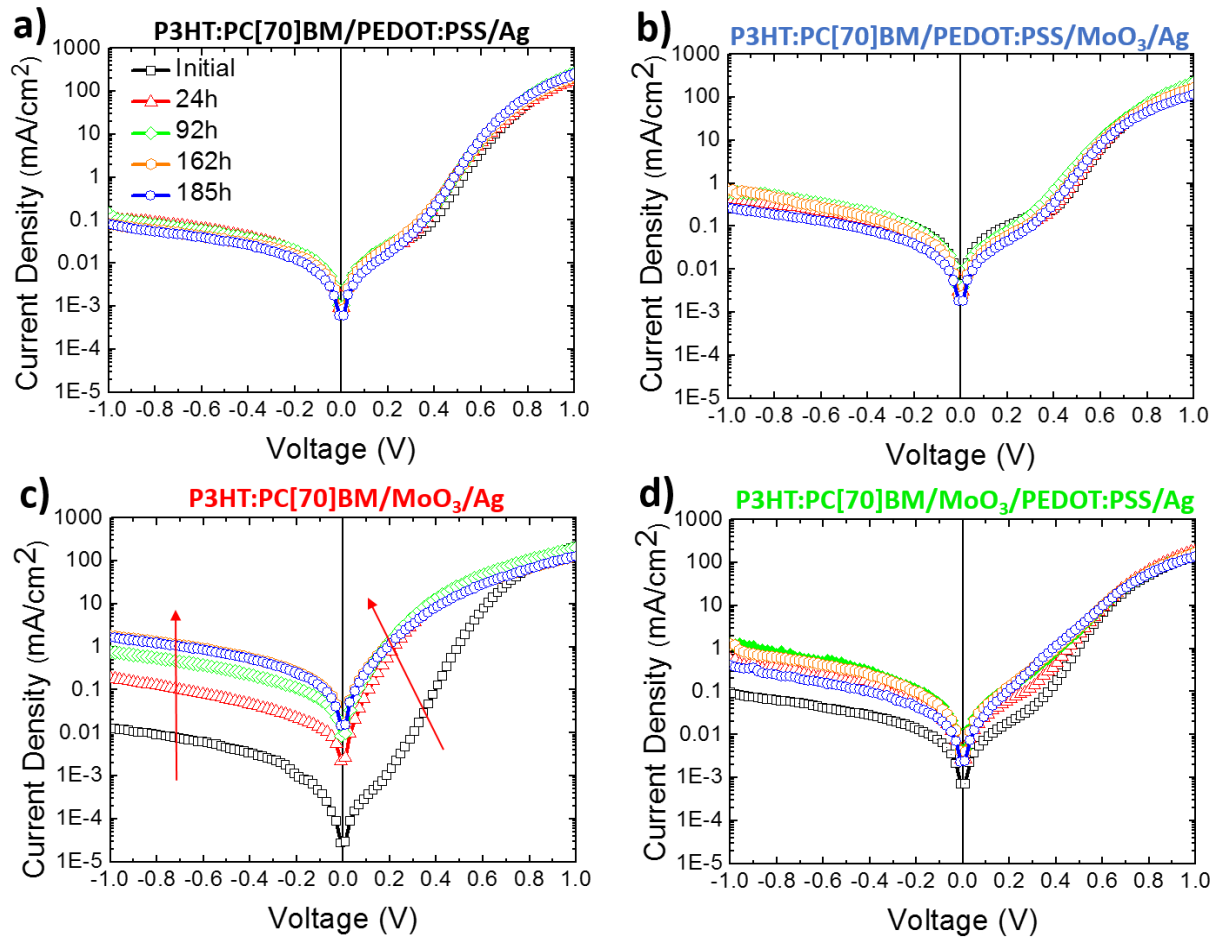

Figure **4**b shows the dark J/V for devices with PEDOT:PSS:ZD/MoO₃ as HTL. The Rp of these devices is slightly improved over time whereas Rs is slightly increased over time. We assume that the Rp is improved due to changes in the organic/organic interface. However, we attribute the slight reduction in Rs to the interactions of the double interlayers affecting the carrier transportation in the vertical direction.

Inverted OPVs using MoO₃ as HTL show slightly reduced Rs over time of exposure .On the other hand, a significant leakage current and ideality factor increase over time, implying poor diode properties and MoO$_3$/active layer interactions (

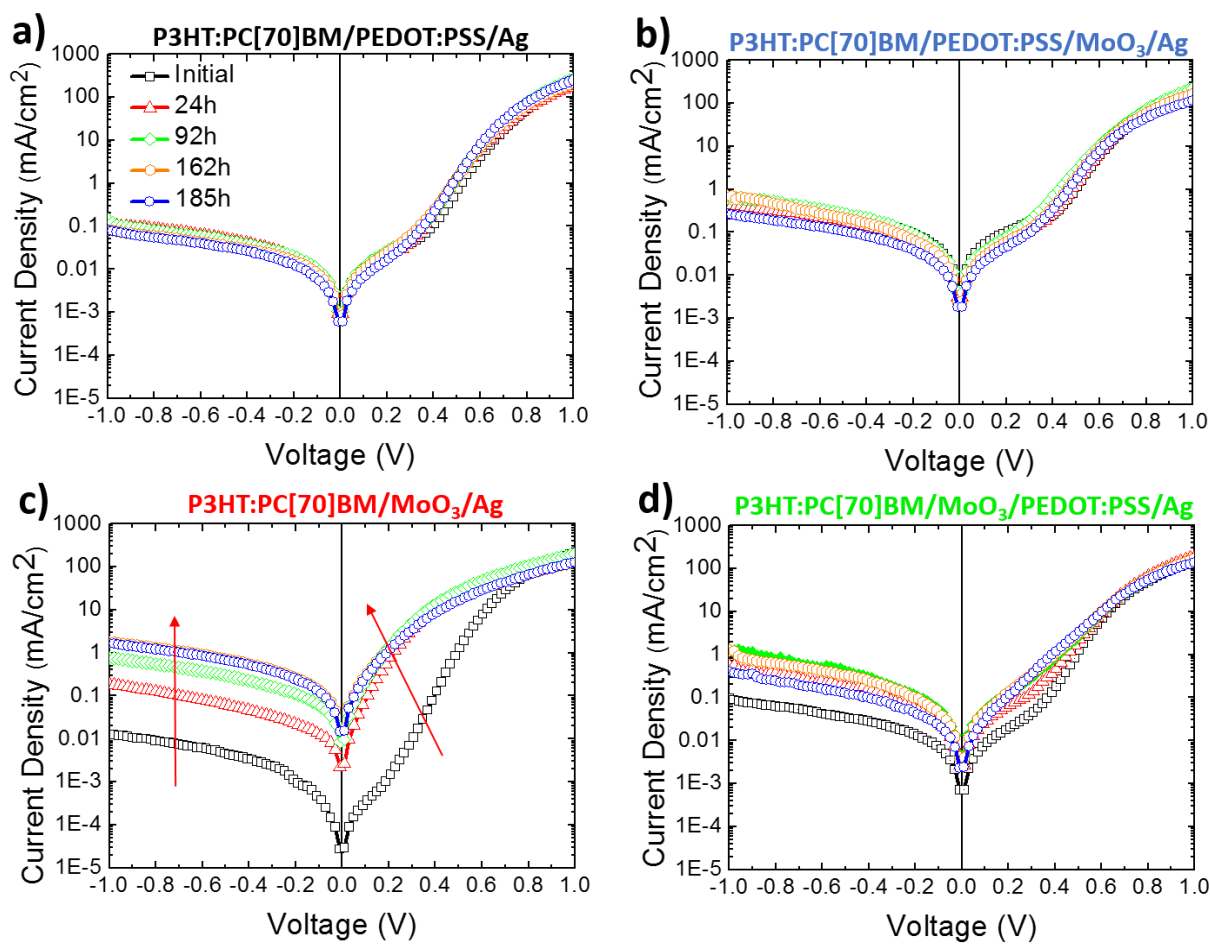

Figure **4**c). The increase in Rp of those devices is more intense and more obvious than the other inverted OPVs under study and it occurs even after some hours of degradation. In addition, the Rp and ideality factor issues over time of exposure are also observed when inverted OPVs using MoO$_3$/PEDOT:PSS:ZD as HTL are tested (Figure 4d).

The latter reveals the detrimental influence of the P3HT:PC[70]BM/MoO$_3$ interface in carrier selectivity.[42] This interfacial interaction over time of exposure promotes high leakage current resulting in poor Rp and ideality factors which are reflected in FF, as also shown in Figure 3.

### 3.3. Investigating the degradation mechanisms of P3HT:PC[70]BM/MoO₃-based inverted OPVs under the ISOS-D-2 protocol using reverse engineering.

Having seen that the inverted OPVs in which MoO₃ directly interfaces with the active layer degrade the fastest, we wanted to investigate whether this interfacial interaction was reversible. We therefore fabricated incomplete device structures with ITO/ZnO/P3HT:PC[70]BM and ITO/ZnO/P3HT:PC[70]BM/MoO₃ stacks. These were aged at 65 °C prior to the deposition of the subsequent fresh MoO₃/Ag and Ag layers, respectively.

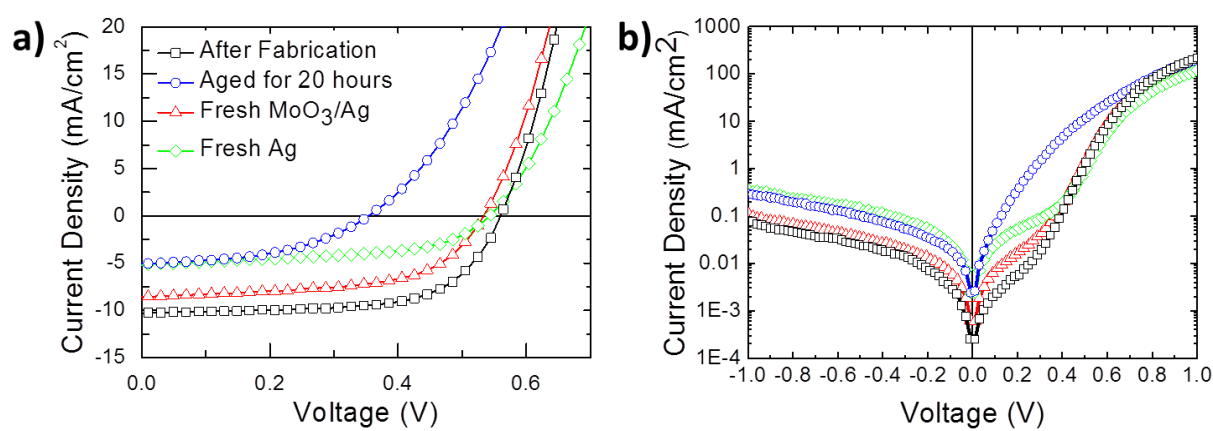

**Figure 5.** (a) Illuminated and (b) dark J/V characteristics of complete devices with MoO₃ HTL as produced (black rectangles) and aged for 20 hours (blue circles). The incomplete stacks were aged for 20 hours at 65 °C and then coated with the required fresh electrode. The stacks were ITO/ZnO/P3HT:PC[70]BM, which was coated with fresh MoO₃/Ag (red triangles) and ITO/ZnO/P3HT:PC[70]BM/MoO₃, which was coated with fresh Ag (green diamonds).

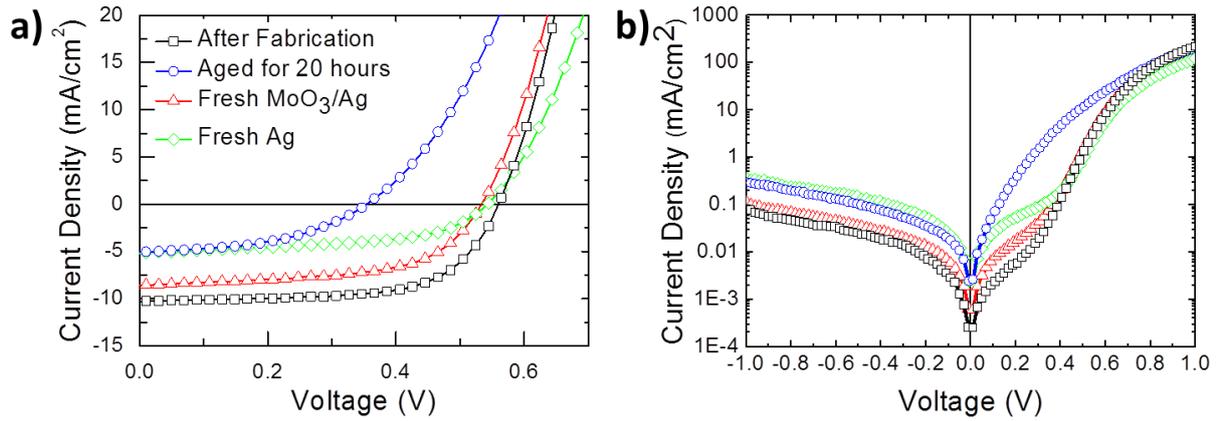

Figure  illustrates the illuminated and dark J/V curves for devices with MoO$_3$ HTL as produced and after aging for 20 hours at 65 °C. Also shown are the J/V characteristics for devices with degraded stacks after fresh evaporation of the top contact, enhancing the aspect that the MoO$_3$/Ag interface also plays a role in this degradation. After reverse engineering, ,[9] illuminated J/Vs show the Voc value is recovered when a new MoO$_3$/Ag is deposited and also when only a new Ag layer is deposited on the ITO/ZnO/P3HT:PC[70]BM/MoO$_3$ aged stack. This seems to suggest that the MoO$_3$/Ag interface is responsible for the Voc decay in devices using MoO$_3$/Ag top contact.

In addition, the rectification of devices with an active layer/MoO₃ interface changes over time and is getting more symmetrical (

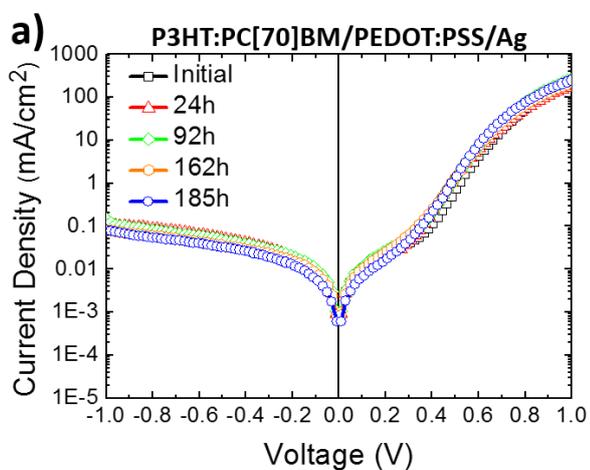

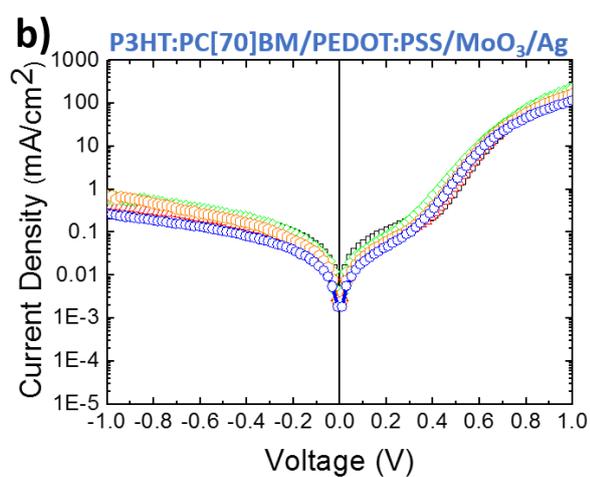

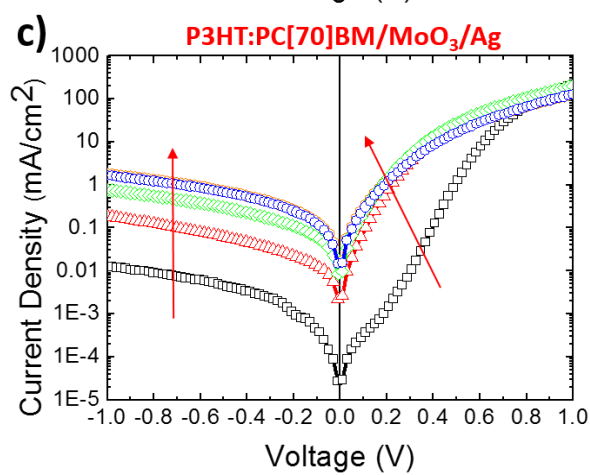

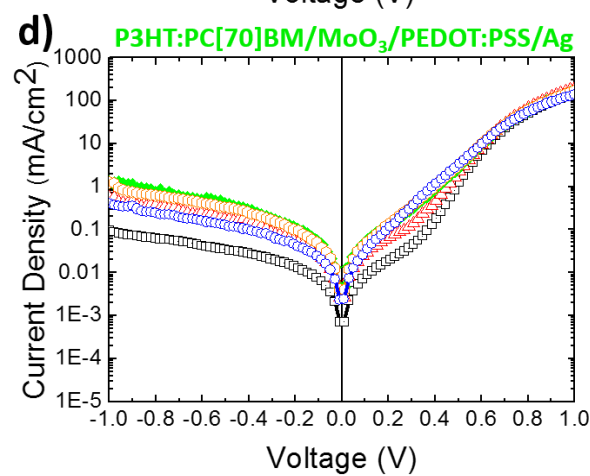

Figure **4**c,

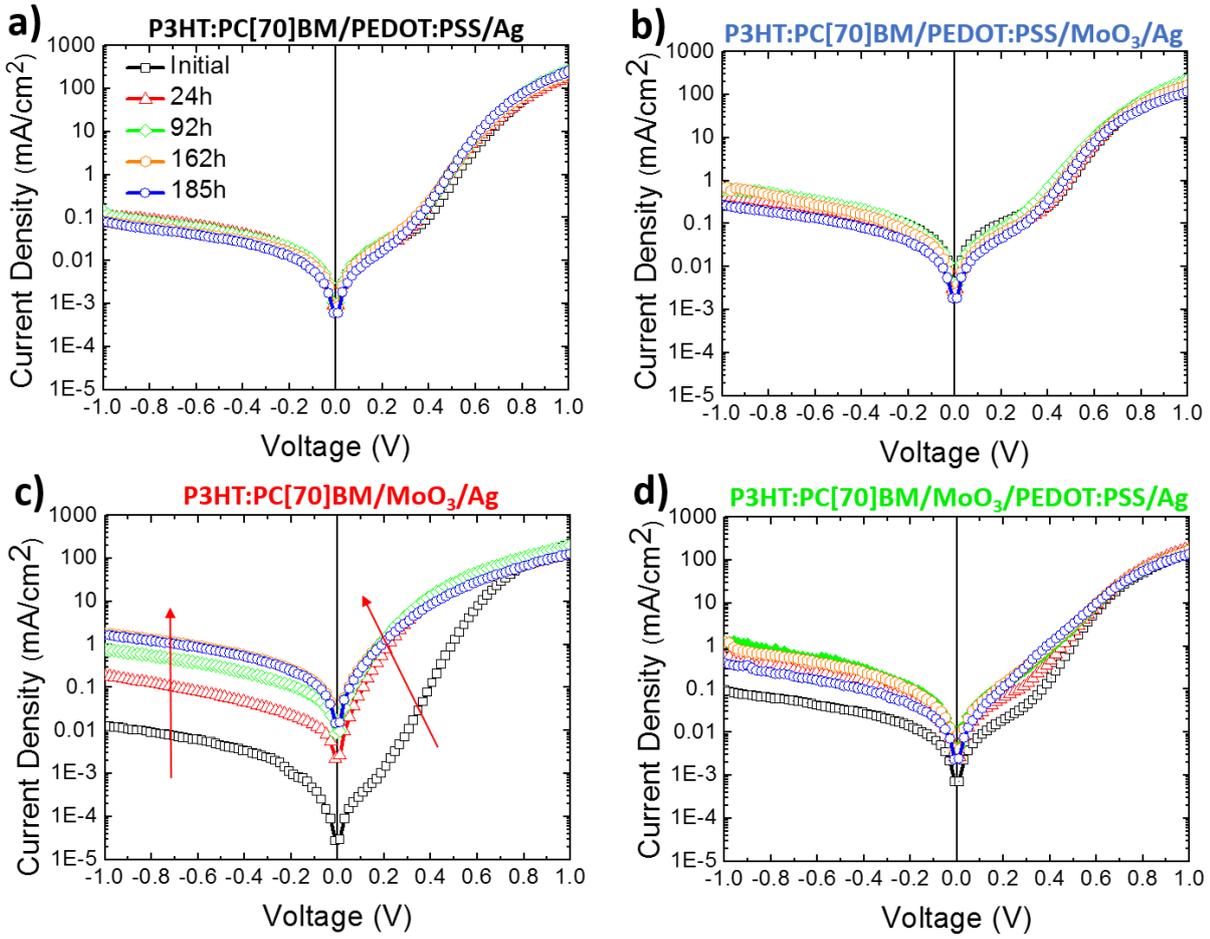

Figure **4**d and

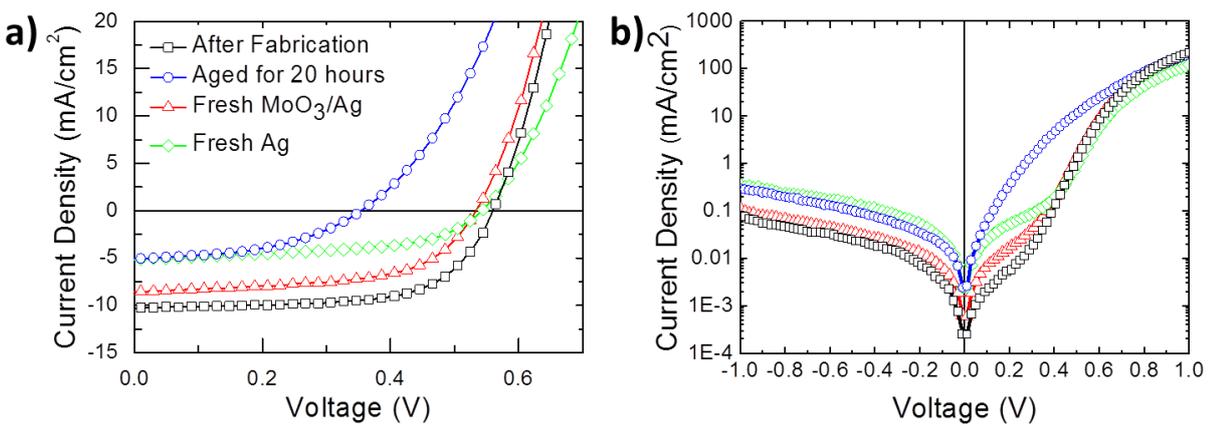

Figure b). However, rectification of complete devices based on aged

ITO/ZnO/P3HT:PC70BM/MoO₃ upon evaporation of a fresh silver electrode, or aged

ITO/ZnO/P3HT:PC70BM with fresh MoO₃/Ag the ideality factor of devices with ITO/ZnO/P3HT:PC70BM/MoO₃ and ITO/ZnO/P3HT:PC70BM aged stacks is recovered (see Figure 5) and is similar in shape to the as-produced devices. This is an indicator that in these devices aging of the MoO₃/Ag interface is responsible for the change in ideality factor.[38]

However, the fresh evaporation of Ag by itself does not lead to a recovery of Jsc. This happens only upon fresh evaporation of both MoO₃ and Ag on the aged stack of ITO/ZnO/P3HT:PC[70]BM. This could be another indication that P3HT:PCBM/MoO₃ interface is responsible for the degraded Jsc of the device over time of exposure under heat conditions. Overall, with a freshly evaporated MoO₃/Ag electrode, we recover 90% of the efficiency of the as-produced devices, which is a clear indication that the device degradation is not affected by ITO/ZnO/P3HT:PCBM interfaces but almost exclusively by the interfaces of the top electrode P3HT:PCBM/MoO3/Ag as shown in

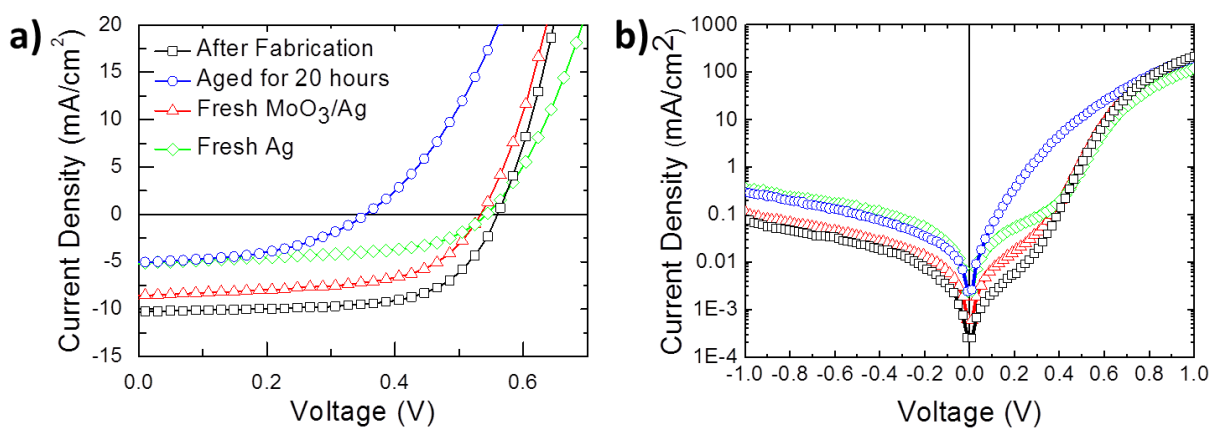

Figure . It should be noted that in principle we expected to reach the same efficiency with the as-produced devices. However, we assume that the exposure to moisture in combination with the heat causes faster degradation and interface modification than the uncapped devices, thereby preventing full recovery by reverse engineering the top contact.

## 4. Summary

We monitored the lifetime performance of P3HT:PC[70]BM, PTB7:PC[70]BM and DPPTTT:PC[70]BM inverted OPVs using thermally evaporated MoO₃ as HTL under heat conditions according to ISOS-D-2 protocol (T = 65 °C, dark, RH = ambient constant at ~40%). Inverted P3HT:PC[70]BM OPVs using thermally evaporated $MoO_3$ as HTL presented the most dramatic decay of all OPV parameters under 65 °C even after a few hours of exposure. A series of experiments were performed using photocurrent mapping and different device architectures in order to isolate the interfaces and reveal the failure mechanisms of P3HT:PC[70]BM devices under accelerated heat conditions.

Upon incorporation of the two compared HTL (evaporated $MoO_3$ and solution-processed PEDOT:PSS:ZD) in different configurations in inverted OPVs with the P3HT:PC[70]BM active layer, we observed that the P3HT:PC[70]BM/$MoO_3$ interface is the most sensitive part of this degradation triggered at 65 °C. This is primarily shown by the reduction in FF and slower degradation of inverted OPVs with $MoO_3$/PEDOT:PSS:ZD double interlayer, which isolates the $MoO_3$ from the Ag. This was in agreement with the dark J/Vs of inverted OPVs, which show the ideality factor of OPVs with the P3HT:PC[70]BM/$MoO_3$ interface is affected, thereby further revealing the detrimental influence of this interface. After alternative reverse engineering method we showed that the Voc and ideality factor of such devices is recovered when fresh Ag is deposited on top of ITO/ZnO/P3HT:PC[70]BM/$MoO_3$ aged stacks. This showed that the Ag electrode influences the stability of inverted OPVs with $MoO_3$ as HTL under heat conditions and was confirmed by the suppressed decay of inverted OPVs using $MoO_3$/PEDOT:PSS:ZD as HTL.

To conclude, in the case of inverted P3HT:PC[70]BM OPVs with $MoO_3$/Ag top contact, which were extensively studied in this paper by using a series of measurements and device/reverse engineering methods, the results presented indicate that inverted OPV heat degradation is not affected by the ITO/ZnO/P3HT:PCBM interfaces but almost exclusively by the interfaces of the top electrode P3HT:PCBM/$MoO3$/Ag. Although the interface between $MoO_3$ and Ag contributes to degradation, the interface between the P3HT:PC[70]BM active layer and $MoO_3$ is the main origin of failure of inverted OPVs under intense heat conditions. In this study of inverted OPVs with thiophene-based active layer materials, the PEDOT:PSS:ZD HTL resulted in more stable inverted OPV performance compared to $MoO_3$ HTL under ISOS-D-2 heat protocol.

AUTHOR INFORMATION

**Corresponding Author**

*Email: stelios.choulis@cut.ac.cy.

**Author Contributions**

The manuscript was written through contributions of all authors. All authors have given approval to the final version of the manuscript.

ACKNOWLEDGMENT

Funding from the European Research Council (H2020-ERC-2014-GoG project number 647311) is gratefully acknowledged.